\documentclass[a4paper,10pt,twoside]{cpc-hepnp}

\usepackage{multicol}
\usepackage{graphicx}
\usepackage{booktabs}
\usepackage{amssymb,bm,mathrsfs,bbm,amscd}
\usepackage[tbtags]{amsmath}
\usepackage{lastpage}
\usepackage{amsmath}

\begin{document}

\fancyhead[c]{\small }
\fancyfoot[C]{\small }

\footnotetext[0]{}

\title{Have we found conclusive evidence for dark matter through direct detection experiments?}

\author{%
      J. Yang\email{j.yang@sjtu.edu.cn}%
\quad
}
\maketitle

\address{%
 Department of Physics, Shanghai Jiao Tong University , Shanghai 200240, China\\
}

\begin{abstract}
 We employ a SUSY-model-independent method to examine the remaining evidence for the low mass dark matter. Using the XENON100's recent result of 224.6 live days $\times$ 34kg exposure and PICASSO's result published in 2009, we have obtained a constrain of couplings, $\left| {{a_n}} \right| < 0.6$ and $\left| {{a_p}} \right| < 1.0$, corresponding to the spin-dependent cross section of  ${\sigma _n}  < 5.6 \times {10^{ - 38}}$cm$^{2}$ and ${\sigma _p} < 1.6 \times {10^{ - 37}}$cm$^{2}$. Spin-independent isospin violating dark matter model also failed to reconcile the recent result from XENON100 with the positive results from DAMA and CoGeNT.
\end{abstract}



\footnotetext[0]{\hspace*{-3mm}\raisebox{0.3ex}{$\scriptstyle\copyright$}
                }%

\begin{multicols}{2}

\section{Introduction}
Since the first time the term ``dark matter" was proposed by Fritz Zwicky in 1933\cite{Zwicky}, a variety of astrophysical and cosmological observations have provided convincing evidence, indicating that something invisible, however, having great influence on gravity does exist in our universe\cite{Evid1,Evid2,Evid3,Evid4}.
Now, it is generally accepted by astrophysicist that the ordinary atomic matter only makes a contribution of 5\% to the universe, while the dark matter (DM) takes the percentage of 23\% and the remaining 72\% is dark energy\cite{Component}.
Meanwhile, a well-motivated extension of Standard Model, supersymmetric (SUSY) theory, provided excellent candidate for DM particles. It's now the most popular Weakly Interacting Massive Particles (WIMPs) among other non-standard WIMP models. \\

In light of WIMP's property that it has weakly or less-than-weakly interaction with ordinary matter and the precondition, supported as well by a recent observation\cite{Evid5}, that the Milky Way is embraced in the dark halo, DM can be searched for directly on earth. The terrestrial experiments designed for DM direct detections are all based on these preconditions. The energy deposited on the target after each hit by a passing DM particle will be transferred to detectable signals such as ionization, scintillation and phonon.\\

The two modes of WIMP-nucleon interaction, which the direct detectors are aiming at, are spin-independent (SI) and spin-dependent (SD). SI coupling, the scalar mode, describes coherent interactions of the entire nucleus with WIMP. SD coupling, the axial one, describes the interaction depending on spin-content of the nucleus. Even though one of these experiments, DAMA/LIBRA, has been claiming success in finding signals of DM for more than a decade\cite{DAMA1,DAMA2}, other experiments only give out null result and thus have made exclusion in the cross section and mass space for SI interaction\cite{CDMS,XENON10}. Recent reports from CoGeNT and CRESST shows as well an evidence\cite{CRESST,CoGeNT}. A survey\cite{Kelso} presenting consistence for these experiments in low mass area (10-20 GeV/$c^{2}$) in the SI mode drew a great deal of attention. The result of the survey applying the isospin-violating DM (IVDM) -nucleus interaction model, which assumes a different interaction strength between proton and neutron,  however, should be checked by XENON100's updated results\cite{XENON100}.\\

In this article, a method of the calculation for SI WIMP-nucleon scattering will be briefly introduced in Sec.2, which will be done by focusing on the IVDM model. In Sec.3 we focus on the SD interaction to show the up-to-date constrains for coupling constants in the WIMP mass of 10 GeV/$c^{2}$ by analysing several leading experiments. Following that, we present the discussion and conclusion in Sec.4.

\section{Spin-independent interaction}

For elastic WIMP-nucleon scattering, the event rate of an Earth-bound detector can be written as \cite{Lewin}
\begin{equation}
R = {N_T}\frac{{{\rho _D}}}{{{M_D}}}\int {d{E_R}} \int\limits_{{v_{\min }}}^{{v_{esc}}} {\frac{{d\sigma }}{{d{E_R}}}} v \cdot f(\overrightarrow v ,\overrightarrow {{v_E}} ){d^3}v
\end{equation}
where $N_{T}$ is the number of target nuclei, the local WIMPs density, $\rho_{D}$=0.3\ GeV/cm$^{3}$\cite{Param1}, divided by the WIMP mass, $M_{D}$, denotes the local number density of WIMPs. The integral interval of recoil energy $E_{R}$ are determined by experimental considerations. The lower limit of the velocity integral, ${v_{\min }} = \sqrt {{M_A}{E_R}/2\mu _A^2}$,  is the minimal value for a WIMP particle to deposit energy $E_{R}$ on to a target atom, whose mass is $M_{A}$, and $\mu_{A}$ is the WIMP-nuclei reduced mass. The upper limit $v_{esc}=544\pm64$ km/s\cite{Param1} is the local Galactic escape velocity. Maxwellian distribution, with characteristic velocity $v_{0}=220$ km/s\cite{Param1}, for $f(v)$ is assumed for a staple dark halo model\cite{Lewin}, where $\overrightarrow v$ is the DM velocity onto the detector while $\overrightarrow {v_{E}}$ is the earth velocity relative to the static galaxy.\\

The differential cross section, $d\sigma /d{E_R}$, is model-dependent, which can be written in a general form,
\begin{equation}\label{eq2}
\frac{{d\sigma }}{{d{E_R}}} = \frac{{\sigma_A{M_A}}}{{2{v^2}\mu _A^2}}{F^2}({q^2})
\end{equation}
where the zero-momentum-transfer cross section ${\sigma_A}$ and the form factor $F^2({q^2})$ are different for the two interaction modes.
\\

In SI mode\cite{IVDM}, the form factor proposed by Helm\cite{Helm} is applicable to various sorts or targets, and the general form of cross section can be
\begin{equation}
\sigma _A^{SI} = \frac{4}{\pi }\mu _A^2{[{f_p}Z + {f_n}(A - Z)]^2}
\end{equation}
where ${f_p}$ and ${f_n}$ are the coupling constants for WIMP-proton and -neutron scattering, which for Majorana WIMPs is usually taken for equality, ${f_p} \simeq {f_n} = f$. Thus Eq. (3) turns to $\sigma _A^{SI} = (4{f^2}/\pi )\mu _A^2{A^2}$, that is proportion to the square of nucleon number $A$. Otherwise, ${f_p} \ne {f_n}$ is called the isospin-violating Dark Matter (IVDM). In Ref.\cite{IVDM}, the assumption of a rate, $f_{p}/f_{n}=-0.7$, succeeded to break away from the constrain of XENON100 and reconciled CoGeNT with DAMA. Here, we apply the latest result of XENON100 with the same IVDM coupling to make a exclusion curve and check the agreeable region of CoGeNT with DAMA in Ref.\cite{IVDM}, finding that the once successful model failed the exam this time, shown in Fig.1. As XENON100 uses xenon as target,which has seven isotopes, the set of $f_{p}/f_{n}=-0.7$ has reached its limit to reduce the constrains from the detector. It will not be able to invalidate XENON100's boundary any longer.
\begin{center}
\includegraphics[width= 8.7cm]{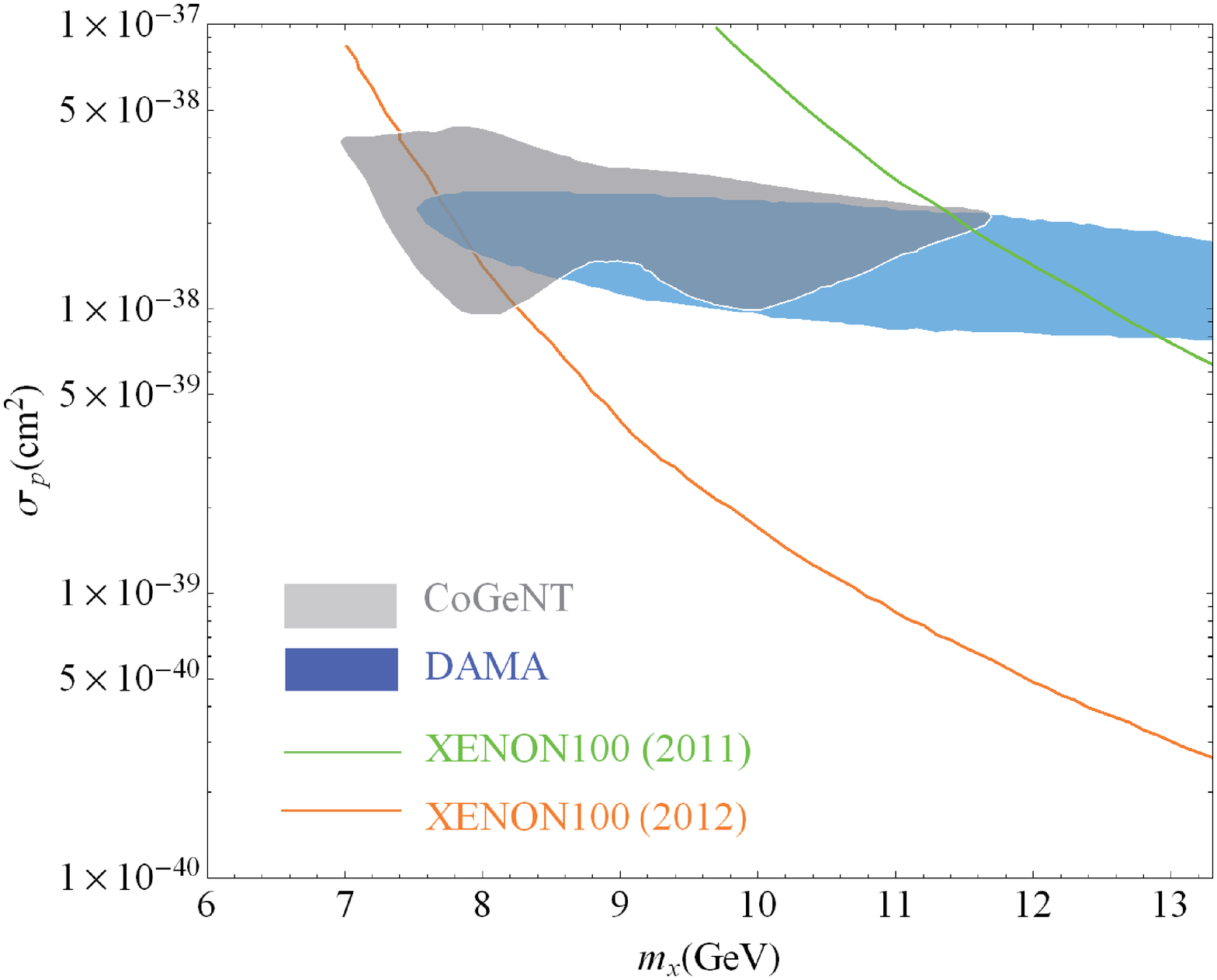}
\figcaption{\label{fig1}Favored region by CoGeNT/DAMA and the exclusive curve by XENON100 for IVDM with $f_{p}/f_{n}=-0.7$. }.
\end{center}

However, the set of $f_{p}/f_{n}=-0.7$ is specific. A more general method of deciding the cross section (the coupling constants) for WIMP's SI scattering provides an overall picture for a given WIMP mass, which is shown in Fig.2. The up-to-date $f_{n}-f_{p}$ region is constrained by XENON100's recent result as an orange ellipse.

\begin{center}
\includegraphics[width= 8.7cm]{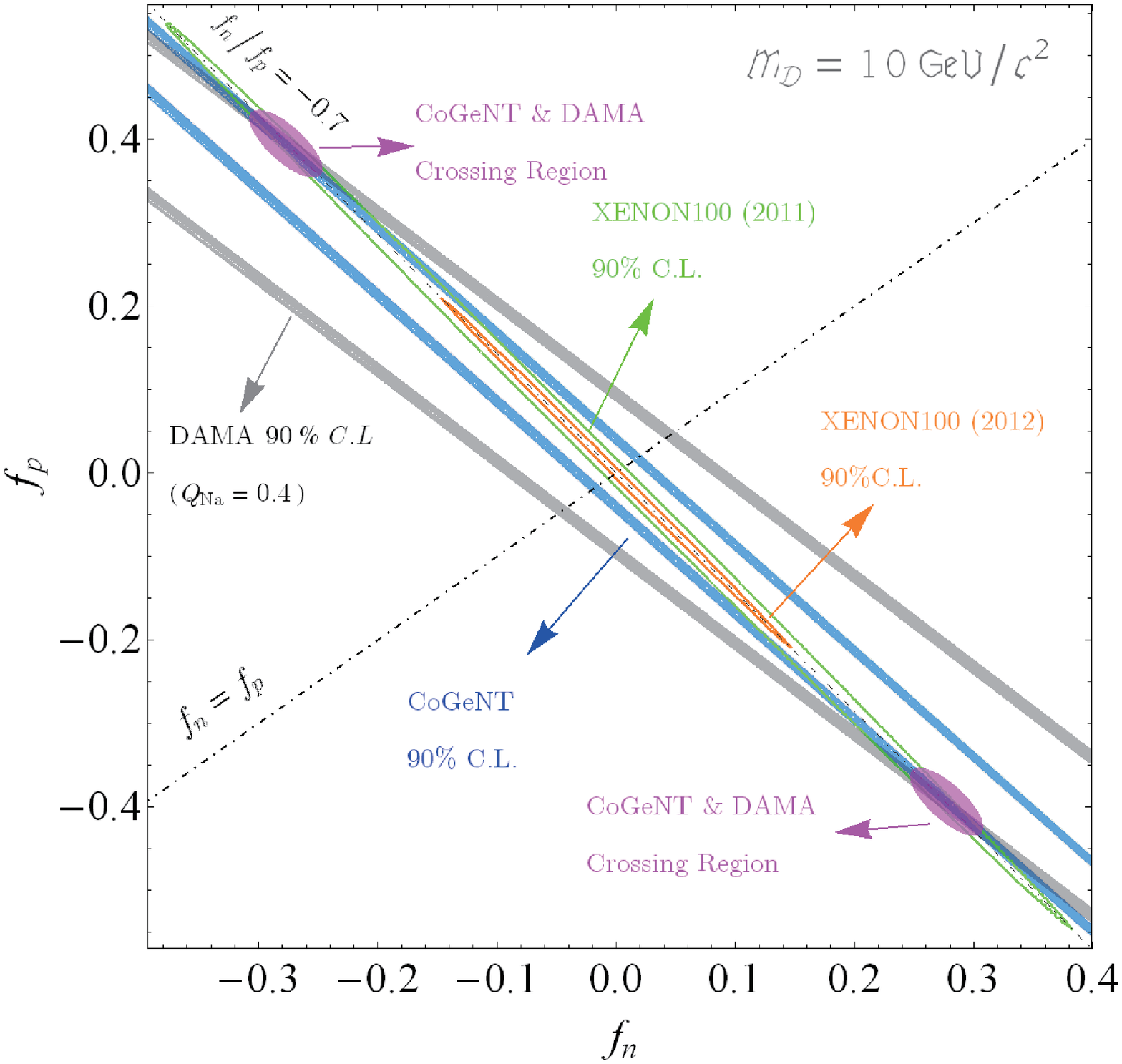}
\figcaption{\label{fig2}Spin-independent couplings for a 10 GeV/$c^{2}$ WIMP}.
\end{center}

In Fig.2 the allowed coupling constants, $f_{p}$ and $f_{n}$, is the inner region of the ellipses. Colored bands represent the allowed parameters from the CoGeNT and DAMA evidence. Although XENON100(2011)'s results, the green ellipse, left some room for CoGeNT and DAMA's evidence, shown in the purple regions, the recent XENON100's results, the orange one, has completely excluded them.
\section{Spin-dependent interaction}
Besides the well-motivated Majorana fermion, a superposition of bino, wino and higgsino appears to be recently the most popular model for WIMP candidate. The Majorana WIMP's interaction can be both SI, introduced in Sec.2, and SD, to be discussed in this Section. The gauginos-higgsinos mixed WIMP(or the so called the lightest SUSY particle, LSP)-nucleon interaction is relevant to its components. Since the precise composition of the LSP is unknown, it is preferable to use a SUSY-model-independent method\cite{Tovey} to do the survey.\\

The general form of SD cross section is given as\cite{SS}
\begin{equation}\label{eq4}
\sigma _{A}^{SD}(q) = \frac{{32G_F^2\mu _A^2}}{{2J + 1}}S(q)
\end{equation}
where $G_{F}$ is the Fermi constant,$\mu_{A}$ is the nucleus-WIMP reduced matter. When normalized, $S(q)/S(0)$ is the form factor for SD mode, which is the counterpart of $F^{2}(q^{2})$ in SI mode. The expression of $S(q)$ can be expanded to
\begin{equation}
S(q) = [a_p^2{S_{pp}}(q) + {a_p}{a_n}{S_{pn}}(q) + a_n^2{S_{nn}}(q)]
\end{equation}
\begin{align}
\nonumber{S_{pp}} &= {S_{00}}(q) + {S_{01}}(q) + {S_{11}}(q) \\
{S_{nn}} &= {S_{00}}(q) - {S_{01}}(q) + {S_{11}}(q) \\
\nonumber{S_{pn}} &= 2[{S_{00}}(q) - {S_{11}}(q)]
\end{align}
$S_{i,j}$ ($i,j=p,n$) is the spin structure, that is specific to each nucleus. The coupling constants are denoted as $a_{p}$ and $a_{n}$. In the case of zero momentum transfer,
\begin{equation}\label{eq7}
S(0) = \frac{{2J + 1}}{\pi }{\lambda ^2}J(J + 1)
\end{equation}
where $\lambda$ is given in the form\cite{J.Engel},
\begin{equation}\label{eq8}
\lambda  = \frac{{\left\langle {N\left| {({a_p}{S_p} + {a_n}{S_n})\left| N \right.} \right.} \right\rangle }}{{\left\langle {{N\left| {\hat J} \right.}}
 \mathrel{\left | {\vphantom {{N\left| {\hat J} \right.} N}}
 \right. \kern-\nulldelimiterspace}
 {N} \right\rangle }} = \frac{1}{J}\left( {{a_p}\left\langle {{S_p}} \right\rangle  + {a_n}\left\langle {{S_n}} \right\rangle } \right)
\end{equation}
From Eq.~(\ref{eq4})(\ref{eq7})(\ref{eq8}), the zero momentum cross section for SD is derived
\begin{equation}\label{eq9}
\sigma _{A}^{SD} = \frac{{32}}{\pi }G_F^2\mu _A^2{[{a_p}\left\langle {{S_p}} \right\rangle  + {a_n}\left\langle {{S_n}} \right\rangle ]^2}\frac{{J + 1}}{J}
\end{equation}
where $\left\langle {{S_p}} \right\rangle$ and $\left\langle {{S_n}} \right\rangle$ are the expectation of spin for proton and neutron by the odd-group nuclear model, $J$ is the total nuclear angular momentum. For single nucleon whose spin and total angular momentum are of the same value, $J = S = \pm1/2$,
\begin{equation}\label{eq10}
{\sigma _{p,n}} = \frac{{24}}{\pi }G_F^2\mu _p^2a_{p,n}^2
\end{equation}
Thus Eq.~(\ref{eq10}) can be expressed by single nucleon cross section
\begin{equation}\label{eq12}
\sigma _A^{SD} = \frac{4}{3}\frac{{\mu _A^2}}{{\mu _p^2}}{\left( {\left\langle {{S_p}} \right\rangle \sqrt {{\sigma _p}}  + \left\langle {{S_n}} \right\rangle \sqrt {{\sigma _n}} } \right)^2}\frac{{J + 1}}{J}
\end{equation}
Owing to the relation of $a_{p}$ and $a_{n}$, it is impossible to work out the single nucleon cross section(or its upper limit) with WIMP. Although we can set a ration of $a_{p}/a_{n}$, it becomes SUSY-model-dependent. Experiments\cite{SIMPPLE,ZEPLIN,PICASSO} choose to set $a_{p}$ to zero when calculating the WIMP-neutron cross section, and vice versa. This is done by assuming $\sigma _A^{} \buildrel \Delta \over = \sigma _A^{n,p}$, where
\begin{align}\label{eq12}
\nonumber\sigma _A^p = \frac{{32}}{\pi }G_F^2\mu _A^2{\left( {{a_p}\left\langle {{S_p}} \right\rangle } \right)^2}\frac{{J + 1}}{J}\\
\sigma _A^n = \frac{{32}}{\pi }G_F^2\mu _A^2{\left( {{a_n}\left\langle {{S_n}} \right\rangle } \right)^2}\frac{{J + 1}}{J}
\end{align}
Despite the ambiguous physical meaning, the $\sigma _A^{p}$ and $\sigma _A^{n}$ are two useful auxiliary mathematical expressions, with the aid of which, the SD cross section can be rewritten as
\begin{equation}\label{eq13}
{\sigma _A} = {(\sqrt {\sigma _A^p}  \pm \sqrt {\sigma _A^n} )^2}
\end{equation}
the sign of which is identical with the sign of $\left\langle {{S_p}} \right\rangle /\left\langle {{S_n}} \right\rangle $.
The comparison between Eq.~(\ref{eq10})and Eq.~(\ref{eq12}) gives
\begin{equation}\label{eq14}
\sigma _{n,p}^{} = \frac{1}{{{{\left\langle {{S_{n,p}}} \right\rangle }^2}}}\frac{J}{{J + 1}}\frac{3}{4}\frac{{\mu _p^2}}{{\mu _A^2}}\sigma _A^{n,p}
\end{equation}
After combining Eq.~(\ref{eq2})(\ref{eq4})(\ref{eq9}), we find the integration of Eq.~(1) brings out an elliptical curve in the $a_{p}$-$a_{n}$ coordinate system, which can be expressed as
\begin{equation}\label{eq15}
N = Aa_p^2 + B{a_p}{a_n} + Ca_n^2
\end{equation}
Eq.~(\ref{eq15}) is a conic section, which can be an ellipse, whose center is the origin of coordinates, if ${B^2} < 4AC$. Otherwise it can be two open curves: hyperbola when ${B^2} > 4AC$ or two parallel lines if $B =  \pm 2\sqrt {AC} $. If the target contains only single nucleon that is sensitive to SD interaction, where being sensitive for SD interaction means the nucleon should be odd-neutron or odd-proton to have non-zero spin, the Eq.~(\ref{eq15}) changes to
\begin{equation}\label{eq16}
N' = {(A'{a_p} + C'{a_n})^2}
\end{equation}
which shows that the ellipse has degenerated into two parallel lines.
If Eq.~(\ref{eq12}) is used to calculate the pure neutron and pure proton cross section (or their upper limits), actually, it's taking the two elliptical points on the coordinate axes, ($a_{n}$, 0) and (0, $a_{p}$), to obtain the $\sigma_{p}$ and $\sigma_{n}$ varying with the WIMP mass.\\

The XENON100 experiment uses the targets with an isotopes abundance of 26.2\% for ${}^{129}Xe$ and 21.8\% for ${}^{129}Xe$, both of which are sensitive to WIMP-neutron SD scattering. The result shows that two candidate events have been observed in the energy range of 6.6-30.5 keV$_{nr}$ from the 224.6 live days $\times$ 34 kg exposure\cite{XENON100}. In the analysis, a set of energy range of 6.6-43.3 keV$_{nr}$ (3-30PE) is used with the expected background of $1.0\pm0.2$. Applying the Feldman-Cousins procedure\cite{Feldman}, we obtained, up to the present, the strongest limit of $a_{n}$. In the calculation, we apply the spin structure from Menendez, Gazit and Schwenk's work (MGS for short)\cite{MGS} and another work by Ressell and Dean (RD)\cite{RD} is taken for reference.\\

Since evidences from CoGeNT and DAMA matching in the WIMP mass of 10 GeV/$c^{2}$\cite{Kelso} have not been checked in the SD mode, we calculated at this mass point to find out the agreeable region in the $\sigma_{p}$-$\sigma_{n}$ area. This can be done by the method introduced above or use an alternative way\cite{Tovey}: if the nuclear cross section (for evidence found) or its upper limit (for null result) is reported, which is obtained in the procedure that account the total events' rate for a single isotope (Eq.~(\ref{eq9})--(13)), we can use Eq.~(\ref{eq17}) straightly,
\begin{equation}\label{eq17}
\sum\limits_{{A_i}} {{{(\frac{{{a_p}}}{{\sqrt {\sigma _p^{\lim ({A_i})}} }}{\kern 1pt} {\kern 1pt}  \pm {\kern 1pt} {\kern 1pt} {\kern 1pt} {\kern 1pt} \frac{{{a_n}}}{{\sqrt {\sigma _{_n}^{\lim ({A_i})}} }})}^2}}  = \frac{\pi }{{24G_F^2\mu _p^2}}
\end{equation}
where $\sigma _{p,n}^{\lim (Ai)}$ is the proton (neutron) experimental upper limit for a certain nuclide, which for evidence found experiments is the certain cross section.\\

We take PICASSO's\cite{PICASSO} result for $a_{p}$ constrain since its low threshold energy qualifies it to provide an exam in the low mass (10 GeV/$c^{2}$) region. The experiment use the superheated $C_{4}F_{10}$ to capture the droplet's explosion into a vapor bubble after hitting by a WIMP. In virtue of its single nuclear interaction in SD mode, we can easily obtain the $\sigma _n^{\lim A}$ from the published $\sigma _p^{\lim A}$ by
\begin{equation}\label{eq18}
\frac{{\sigma _p^{\lim A}}}{{\sigma _n^{\lim A}}} = \frac{{{{\left\langle {{S_n}} \right\rangle }^2}}}{{{{\left\langle {{S_p}} \right\rangle }^2}}}
\end{equation}
The spin expectation is listed in Tabel~\ref{tab1}.
\begin{center}
\tabcaption{ \label{tab1}  Spin values for relevant nuclides.}
\footnotesize
\begin{tabular*}{80mm}{c@{\extracolsep{\fill}}ccccc}
\toprule Nucleis &   Odd  & J & $\left\langle {{S_p}} \right\rangle $  & $\left\langle {{S_n}} \right\rangle $ & Ref. \\
\hline
${}^{19}F $\hphantom{000}&p& 1/2 & 0.441 & -0.109 & \cite{F} \\
${}^{23}Na$\hphantom{00} &p& 3/2 & 0.248 & 0.020 & \cite{RD} \\
${}^{73}Ge$\hphantom{00} &n& 9/2 & 0.030 & 0.378 & \cite{Ge} \\
${}^{129}Xe$\hphantom{000} &n& 1/2 & 0.010 & 0.329 & \cite{MGS} \\
${}^{131}Xe$\hphantom{000} &n& 3/2 & -0.009 & -0.272 & \cite{MGS} \\
\bottomrule
\end{tabular*}
\vspace{0mm}
\end{center}
In the procedure of dealing with CoGeNT's spectrum\cite{Kelso}, we adopt the most stringent surface event rejection to check the lower limit of its couplings (or cross section) with an $99\%$ confidence level. Other couplings' region under a milder surface rejection will surly be excluded if the most stringent one can even not survive.\\

\begin{center}
\includegraphics[width=8.7cm]{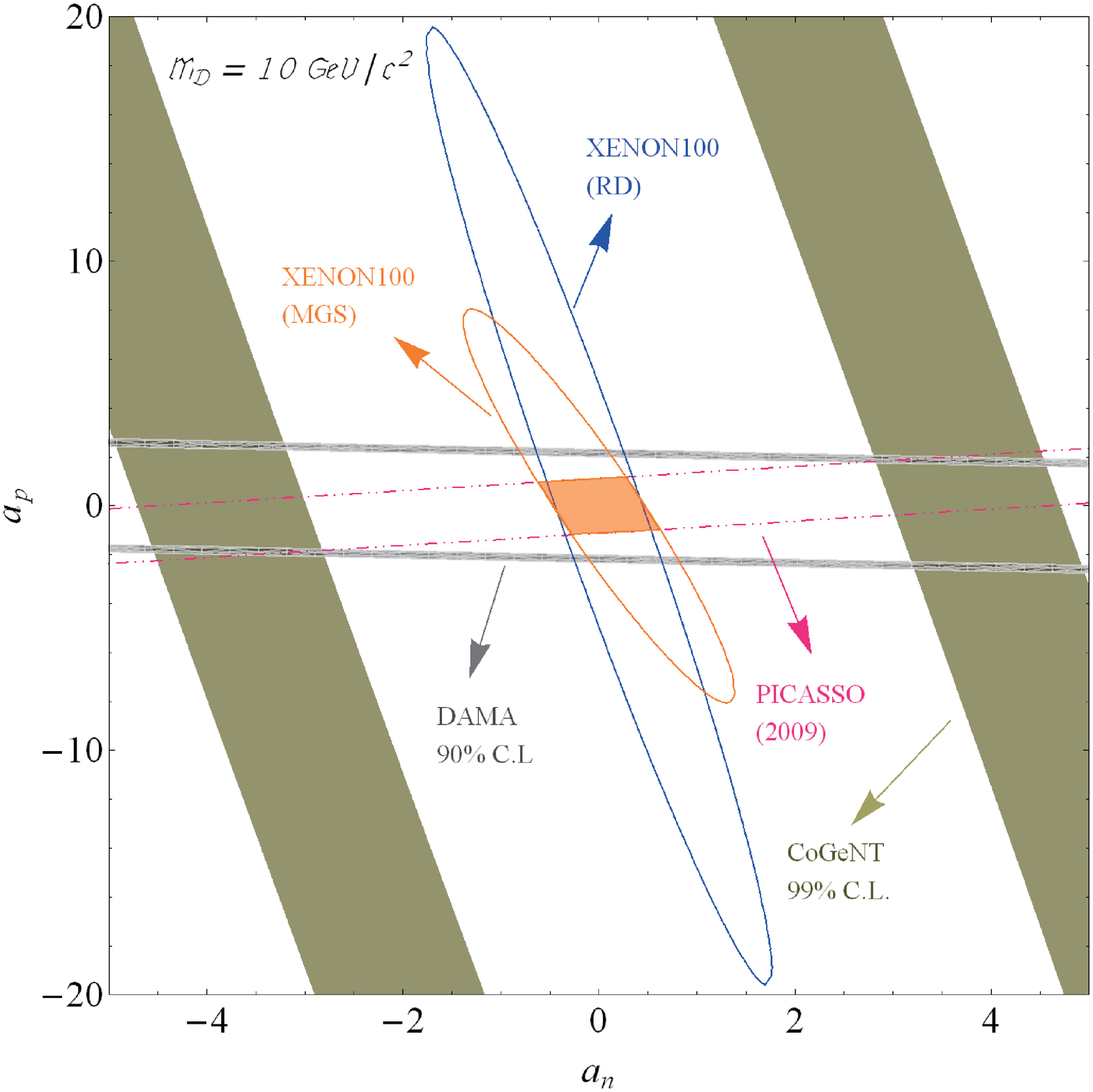}
\figcaption{\label{fig2}Spin-dependent couplings allowed by XENON100 and PICASSO at a WIMP mass of 10 GeV/$c^{2}$ (filled in orange). }.
\end{center}

In Fig.3, the allowed region of $a_{p}$ and $a_{n}$ is constrained both by XENON100's ellipse and PICASSO's dashed lines, which is filled in orange. CoGeNT and DAMA's evidence(colored bands) is excluded obviously.
\section{Discussion and conclusion}
In this article, we have examined the signal reported by DAMA/NAI and CoGeNT with the recent result from XENON100 and PICASSO experiments. The remained unchecked 10 GeV/$c^{2}$ DM evidence is excluded for both the SI IVDM model and the SD model.

\begin{itemize}
  \item The IVDM model with $f_{p}/f_{n}=-0.7$ is unable to keep the CoGeNT and DAMA's compatible region evading from the XENON100's constrain. That is to say, in SI mode, there's no evidence can be affirmed from DM direct detection experiments either.
  \item With the constrains from XENON100 (2012) and PICASSO (2009), for 10 GeV/$c^{2}$ DM mass, we have obtained the allowed couplings' region of $\left| {{a_n}} \right| < 0.6$ and $\left| {{a_p}} \right| < 1.0$, corresponding to the cross section ${\sigma _n}  < 5.6 \times {10^{ - 38}}$cm$^{2}$ and ${\sigma _p} < 1.6 \times {10^{ - 37}}$cm$^{2}$. Thus, in SD mode, no WIMP signal is compatible within all experimental results.
\end{itemize}

For SD calculation, a model-independent method has been introduced. While experiments are used to report the cross section in the pure neutron or pure proton way, it is insufficient to give a full comparison between different detectors. This has been shown in Sec.3 for a 10 Gev/$c^{2}$ WIMP. Other masses are also excluded using the same method.\\

\acknowledgments{I'd like to thank my supervisor K.Ni for useful discussion and helpful advices given on the improvement of this work.}

\vspace{-1mm}
\centerline{\rule{80mm}{0.1pt}}

\end{multicols}

\clearpage

\end{document}